\documentclass[journal]{IEEEtran}
\usepackage{bm}
\usepackage{amsmath}
\usepackage{subfigure}
\usepackage{algorithm}
\usepackage{algorithmic}
\usepackage{booktabs}
\usepackage{color}
\usepackage{verbatim}
\usepackage{setspace}
\usepackage[breaklinks,colorlinks,linkcolor=black,citecolor=black,urlcolor=black]{hyperref}
\usepackage{multirow}

\newcommand*{\red}{\textcolor{red}}

\newcommand*{\blue}{\textcolor{black}}
\newcommand*{\added}{\textcolor{black}}

\usepackage[]{graphicx}    

\ifCLASSINFOpdf
\else
\fi
\begin{document}
%
\title{Simulated Annealing Based Heuristic for Multiple Agile Satellites Scheduling under Cloud Coverage Uncertainty}
%
%
%

\author{Chao Han, 
        Yi Gu, 
        Guohua Wu,~\IEEEmembership{Member,~IEEE,} 
        and~Xinwei Wang
\thanks{Chao Han and Yi Gu are with the School of Astronautics, Beihang University, Beijing, 100191, China (e-mail:hanchao@buaa.edu.cn; guyi\_buaa@buaa.edu.cn)}
\thanks{Guohua Wu and Xinwei Wang are with School of Traffic and Transportation Engineering, Central South University, Changsha, 410000, China (e-mail:guohuawu@csu.edu.cn; xinwei.wang.china@gmail.com) (\textit{Corresponding author}: Xinwei Wang)}
\thanks{This work has been submitted to the IEEE for possible publication. Copyright may be transferred without notice, after which this version may no longer be accessible.}}

%
%

\markboth{IEEE TRANSACTIONS ON SYSTEMS, MAN, AND CYBERNETICS: SYSTEMS}
{Shell \MakeLowercase{\textit{et al.}}: Bare Demo of IEEEtran.cls for IEEE Journals}
%



\maketitle

\begin{abstract}
\noindent Agile satellites are the new generation of Earth observation satellites (EOSs) with stronger attitude maneuvering capability. Since optical remote sensing instruments equipped on satellites cannot see through the cloud, the cloud coverage has a significant influence on the satellite observation missions. We are the first to address multiple agile EOSs scheduling problem under cloud coverage uncertainty where the objective aims to maximize the entire observation profit. The chance constraint programming model is adopted to describe the uncertainty initially, and the observation profit under cloud coverage uncertainty is then calculated via a sample approximation method. Subsequently, an improved simulated annealing-based heuristic combining a fast insertion strategy is proposed for large-scale observation missions. Finally, extensive experiments are conducted to verify the effectiveness and efficiency of the proposed method. Experimental results show that the improved simulated annealing-based heuristic outperforms other algorithms for the multiple AEOSs scheduling problem under cloud coverage uncertainty, which verifies the efficiency and effectiveness of the proposed algorithm.
\end{abstract}


%
\IEEEpeerreviewmaketitle

\section{Introduction}
%
%
%
%
%
%
Earth observation satellite (EOS), which serves as a space platform orbiting the Earth, utilizes remote sensors to collect images of targets on the Earth surface~\cite{liu2017adaptive}.
With the advantage of space locations, the EOS has been applied to many fields, such as resource exploration, weather prediction and disaster alerts~\cite{chen2019mixed}.

\begin{figure*}[htb]
	\begin{center}
		\includegraphics[width=0.75\textwidth]{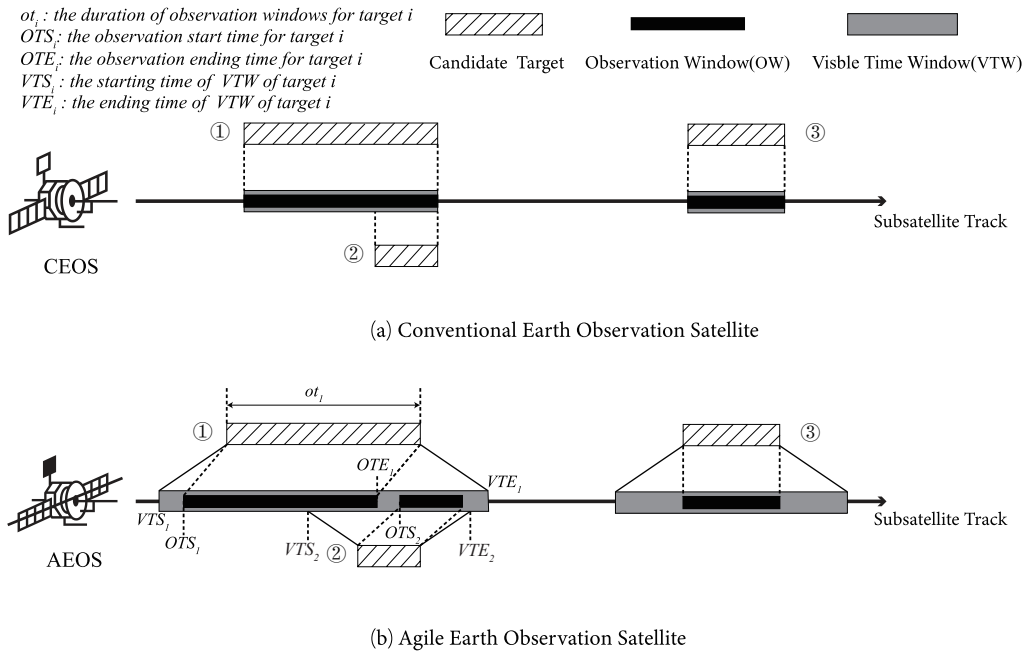}
		\caption{Comparison between CEOS and AEOS.}\label{fig:AEOS}
	\end{center}
\end{figure*}

Agile Earth observation satellites (AEOSs) are the new generation of EOSs. 
Compared to conventional \red{}EOSs (CEOSs), AEOSs have considerable improvement in attitude maneuverability.
Fig.~\ref{fig:AEOS} shows different observation conditions for three candidate targets between CEOS and AEOS, 
in which \red{}visible time window (VTW) is the time window during which the target is visible for \red{}a satellite and \red{}observation window (OW) denotes the actual observation time window.
\red{}Detailed calculation procedure of VTWs can be found in~\cite{wang2019onboard,gu2019kriging}.
With the attitude maneuvering ability of roll axis, CEOS can only perform observation missions when it is right above targets on the Earth surface. 
As shown in Fig.~\ref{fig:AEOS}, the conflict between targets 1 and 2 is therefore inevitable for CEOS owing to the time overlapping. 
However, both two targets can be observed by single AEOS with stronger observation capacity arising from the maneuverability of pitch axis.    
Possessing three degrees of freedom (roll, pitch and yaw axes), AEOS is able to observe targets for longer periods~\cite{HE201812}, which better satisfies users' requirements.

Despite the advantages of AEOS compared to CEOS, the AEOS mission scheduling is more difficult. Through the adjustment of pitch axis, AEOS could access longer VTW for each target, generally exceeding the required observation time~\cite{HE201812}.
Therefore, it is necessary to determine the observation start time for \red{}each target. In Fig.~\ref{fig:AEOS}, the optional range of the \red{}observation start time $OTS_{1}$ for target 1 varies in time period [$VTS_{1}, VTE_1-ot_{1}$], greatly increasing the scheduling complexity.
Meanwhile, the attitude transformation time constraint between two adjacent targets should be taken into consideration; 
the AEOS has to accomplish the roll and pitch attitude transformation between targets 1 and 2. The energy consumption in both attitude maneuvering process and observation operation should also be considered.
Notice that the attitude of AEOS for observation mission depends on the observation start time, which consequently determines the transition time between two adjacent targets. Hence, the selection of \red{}observation start time results in the coupling of feasible transition time constraints and energy constraints.
Given all above characteristics, Lematre $et\ al$.~\cite{LemaitreVerfaillie-269} indicated that the AEOS scheduling problem, although simplified, is NP-hard.

A number of cross-sectional studies contributed to CEOS scheduling problems. 
Gabrel $et\ al$.~\cite{GabrelMoulet-258,GabrelVanderpooten-61} presented a graph-theoretic model and derived exact and inexact algorithms. 
Wolfe and Sorensen~\cite{WolfeSorensen-272} introduced the window-constrained-packing model for EOS scheduling. 
Benoist and Rottembourg~\cite{Benoist2004Upper} provided the upper bound of the observation profit optimization, using the generalized traveling salesman problem with time windows.
Lin $et\ al$.~\cite{LinLiao-60} proposed a Lagrangian relaxation method to solve the daily imaging scheduling of single CEOS.
Marinelli $et\ al$.~\cite{Marinelli2011A} formulated the problem as a time-indexed integer programming. 
Wu $et\ al$.~\cite{wu2013two,wu2015coordinated,wu2017satellite} has taken the mission clustering strategy into satellite scheduling problem and proposed several efficient algorithms.  
Wang $et\ al$.~\cite{wang2018fixed} adopted the flow formulation to describe the fixed interval scheduling of multiple CEOSs.
Considering the feature of limited resource capacities, Chen $et\ al$.~\cite{chen2019mixed} proposed a mixed integer linear programming for multi-satellite scheduling.
\red{}Hu $et\ al$.~\cite{hu2019branch} designed a branch-and-price algorithm for CEOSs constellation imaging and downloading integrated scheduling problem.

Meanwhile, enormous research about agile satellite scheduling has been conducted~\cite{wang2020agile}.
The mission management problem of AEOS was first set out and analyzed by Lematre $et\ al$.~\cite{LemaitreVerfaillie-269}.
Beaumet $et\ al$.~\cite{Beaumet2011Feasibility}, Liu $et\ al$.~\cite{liu2016time} and She $et\ al$.~\cite{she2018onboard} studied the scheduling problem of agile satellite autonomous mission planning. Wang $et\ al$.~\cite{wang2016scheduling,wang2019scheduling} connected the theory of complex networks with AEOS redundant targets scheduling problem, and proposed a fast approximate scheduling algorithm.  
A metaheuristic based on large neighborhood search method was introduced to solve the time-dependent scheduling problem in~\cite{liu2017adaptive,HE201812}.
Cho $et\ al$.~\cite{cho2018optimization} formulated a two-step linear programming model for mission planning of satellite constellation.
Li $et\ al$.~\cite{li2018preference} proposed a preference-based evolutionary algorithm in the multi-objective optimization of agile satellite mission planning.
Du $et\ al$.~\cite{du2018area} introduced a novel algorithm for area targets observation in which the drift angle constraint was considered.
\red{}Peng $et\ al$.~\cite{Guansheng2020An} proposed an exact algorithm to solve single AEOS scheduling problem considering time-dependent profits, where the profit of a target depends on the observation start time.

In practical satellite scheduling, the uncertainty,  which origins from the change of mission priority, weather conditions and satellite resource status, is inevitable. Considerable researchers have studied the scheduling problem for both CEOS and AEOS, in which the uncertainty derived from dynamic missions and emergency responses have been considered~\cite{wang2014dynamic,WangZhu-21,du2018new,ntagiou2018ant,he2019hierarchical}. On account of the widespread application of optical sensors on the EOS, the observation mission is extremely influenced by the uncertainty of cloud coverage~\cite{globus2004comparison,wang2019robust}.
It is reported that around 35\% of the images acquired by the Landsat-7 sensors are blocked by cloud~\cite{JU20081196}. He $et\ al$.~\cite{he2013research} also reported that about 60\% current observations were useless owing to the cloud coverage, which demonstrated the significance considering  cloud coverage in satellite scheduling.
However, limited attention for EOS scheduling considering the cloud coverage is obtained.
Liao $et\ al$.~\cite{liao2007imaging} formulated a stochastic integer programming model to represent the influence of cloud coverage. 
Wang $et\ al$.~\cite{WangDemeulemeester-4} employed a chance constrained programming (CCP) model to describe the stochasticity of cloud, and presented a \red{}branch-and-cut method to solve the problem.
Utilizing scenarios representing the uncertainty of cloud coverage, Valicka $et\ al$.~\cite{valicka2019mixed} introduced novel stochastic mixed integer programming models with the objective of maximizing collection quality when cloud cover uncertainty was considered.
Wang $et\ al$.~\cite{8573152} further demonstrated the efficiency of \red{}the proposed exact and heuristic algorithm in the scheduling of \red{}multiple CEOSs under cloud coverage uncertainty. \red{}Besides, Wang $et\ al$.~\cite{wang2020expectation} developed an expectation model and a sample average approximation model, and comparative experiments have been conducted. A two-phased scheduling framework, taking advantage of hypothetic real-time cloud information, was associated to the AEOS scheduling in~\cite{he2016cloud}. Unfortunately, the cloud coverage is always changing, which is quite difficult to be exactly predicted~\cite{Bensana1999Dealing}. 

Several drawbacks still exist for the reviewed papers. First, despite extensive research of AEOS scheduling, few study has taken \red{}the uncertainty of cloud coverage into consideration, especially for the multi-AEOS scheduling problem. Second, current research about CEOS scheduling cannot be readily applied to AEOS uncertain scheduling, since the mathematical model of CEOS scheduling has difficulty transforming into AEOS model. Moreover, the algorithms proposed for CEOS scheduling with uncertainty are not suitable owing to the enhanced attitude maneuverability of AEOS.

\red{}Motivated by the huge impact of the cloud coverage and the lack of corresponding AEOS scheduling research, we therefore address the multi-AEOS scheduling problem \blue{considering energy and memory constraints} under cloud coverage uncertainty.
The CCP model~\cite{liu2009some} is adopted to describe the uncertainty of cloud coverage, and the observation profit is then calculated via the sample approximation method.
With respect to the characteristics of agile satellites, an optimization subproblem, solved by \blue{a} sequential quadratic
programming (SQP) method, is established to determine actual observation time windows.
To solve the master optimization problem of maximizing the entire observation profit under uncertainty, \red{}an improved simulated annealing (ISA) based heuristic is developed, in which a fast insertion strategy plays an important role in arranging missions. 
Afterwards, we conduct extensive experiments to verify the effectiveness and efficiency of the proposed algorithm in solving multi-AEOS scheduling problem under cloud coverage uncertainty. 

The contributions of this study \added{are therefore} summarized as follows: 
        \begin{itemize}
            \item The multiple AEOSs scheduling problem considering energy and memory constraints under cloud coverage uncertainty is addressed in this work. \added{The introduced uncertainties are mathematically represented with CCP and a sample approximation method.}
            \item Utilizing the concept of time slack, an optimization subproblem of determining the observation start time is established\added{, and we then propose a SQP based fast insertion algorithm, which plays a significant role to ensure efficient update of the scheduling solution}.
            
            \item In order to \added{effectively and efficiently solve the uncertain AEOSs scheduling problem}, \added{we propose} an ISA based heuristic  \added{consisting of two stages: an initial solution is constructed using the selection rules of targets and resources at the first stage; and in the second stage, the ISA combining with the fast insertion algorithm is utilized to iteratively update the scheduling solution.} \added{Extensive experiments are conducted to demonstrate the superior performance of our proposed algorithm.}
        \end{itemize}

The remainder of this paper is organized as follows. In Section~\ref{sec:model}, necessary definitions and problem \red{}descriptions are provided, and the CCP model is established. Section~\ref{sec:method} introduces the sample approximation method to calculate the observation profit, defines the optimization subproblem of determining the \red{}start time of OW and proposes \red{}an ISA based heuristic. A series of experimental results are reported in Section~\ref{sec:comput}. We summarize our work and point out future directions in the last section.

\section{Problem statement}
\label{sec:model}
In this section, the deterministic multi-AEOS scheduling problem is introduced at first, followed by the uncertain model considering cloud coverage uncertainty. 
\subsection{Deterministic AEOS scheduling}
Based on the relative size between ground targets and the view horizon of satellite sensors, the observation objects can be divided into point targets and area targets in EOS scheduling. 
\red{}The area target can be decomposed into multiple point targets, therefore, we only consider the point targets in this study. Different satellite orbits are viewed as different orbital resources.  
The following assumptions and simplifications are further provided:

\begin{enumerate}
	\item Each satellite conducts at most one observation mission in one time;
	\item Each target cannot be observed more than once.
	\item The attitude transformation time, energy and memory consumption are taken into consideration.
	\item The process of data download is not considered.  
\end{enumerate}

\begin{table}[htbp]
	\centering
	\caption{Notations.}
	\footnotesize
	\scalebox{1.0}{
		\begin{tabular}{lp{21.25em}}
			\toprule
			\multicolumn{2}{l}{\textbf{Target}}\\
			$A$   & Set of observation targets, $A = \{ 1,...,\added{|A|}\}$ \added{and $|A|$ is the target set size} \\
			$i,j$ & Target index, $i,j\in A\cup \{ 0,\added{|A|}+1\}$, in which $0,\added{|A|}+1$ are dummy targets \\
			$ot_{i}$ & The observation duration of target $i$, $i\in A$ \\
			$\omega_{i}$ & \blue{Observation} profit of target $i, i\in A$ \\
			\multicolumn{2}{l}{\textbf{Orbit}}\\
			$O$   & Set of orbits, $O = \{ 1,...,\added{|O|}\}$ \added{and $|O|$ is the orbit set size} \\
			$k$   & Orbit index, $k\in O$ \\
			$M_{k}, E_{k}$ & Memory and energy capacity \added{on} orbit $k$, $k\in O$ \\
			$m_{k},e_{k}$ & Memory and energy consumption for unit time of observation \added{on} orbit $k$ \\
			$e_{k}^{\prime}$ & Energy consumption for each unit angle of attitude transformation \added{on} orbit $k$ \\
			\multicolumn{2}{l}{\textbf{Decision variables}}\\
			$x_{ik}$ & \blue{Binary} decision variable. $x_{ik} = 1$ if target $i$ is scheduled to be observed on orbit $k$, otherwise $x_{ik} = 0$\added{, $i\in A, k\in O$}  \\
			$TP_{ik}$ & \blue{Continuous decision variable within} $[0,1]$. $TP_{ik}$ is associated to the observation start time for target $i$ on orbit $k$, $i\in A, k\in O$  \\
			\multicolumn{2}{l}{\textbf{Constraint parameters}}\\
			$b_{ik}$ & $b_{ik} = 1$ if orbit $k$ is available for the observation of target $i$, otherwise $b_{ik} = 0$, $i\in A, k\in O$ \\
			$[OTS_{ik},OTE_{ik}]$ & Observation time window  of target $i$ on orbit $k$, $i\in A, k\in O$ \\
			$[VTS_{ik},VTE_{ik}]$ & Visible time window of target $i$ on orbit $k$, $i\in A, k\in O$ \\
			\added{$Trans(i,j,k)$} & \blue{Attitude transformation} time between targets $i$ and $j$ on orbit $k$, $i,j\in A, k\in O$ \\
			$se_{ij}^{k}$ & Energy consumption in the transformation from $i$ to $j$ on orbit $k$. \added{$se_{ij}^{k}=0$ if $j$ is a dummy target}, $i\in A, \added{j\in A\cup \{ |A|+1 \},}~k\in O$ \\
			$\lambda_{ik}$ & Binary stochastic parameter, $\lambda_{ik} = 1$ denotes that target $i$ is successfully observed on orbit $k$,  and $\lambda_{ik} = 0$ otherwise \\
			$p_{ik}$ & Probability that target $i$ will be successfully observed on orbit $k$, $i\in A, k\in O$ \\
			$W$ & Set of sample scenarios and $|W|$ is the sample size \\
			$w_l$ & A scenario, $w_l \in W$ \\
			$y_l$ & Binary variable. $y_{l} = 1$ if current solution is infeasible in scenario $w_{l}\in W$, otherwise $y_{l} = 0$ \\
			
			\multicolumn{2}{l}{\textbf{Objective}}\\
			$\sum\limits_{i\in{A}} \sum\limits_{k\in{O}} \omega_{i}\cdot x_{ik}$ & The entire observation profit for deterministic scheduling problem\\
			$f$ & The entire observation profit for scheduling problem under cloud coverage uncertainty \\
			\bottomrule
	\end{tabular}}%
	\label{tab:Notation}%
\end{table}%

The notations used in our model are summarized in Table~\ref{tab:Notation}.
\added{Denote $A$ and $O$ as the set of targets and orbits, respectively.
\added{The required observation time is denoted as $ot_{i}$ for each target $i\in A$.}
The observation profit for target $i$ is defined as~$\omega_{i}$.
For each orbit $k\in O$, we define the following parameters: memory capacity $M_{k}$, energy capacity $E_{k}$, memory consumption $m_{k}$ and energy cost $e_{k}$ for unit time during observation, and energy consumption $e_{k}^{\prime}$ for  unit-time angle of attitude transformation.}
\added{Binary decision variable $x_{ik}$ is introduced to represent whether target $i$ is scheduled to be observed on orbit $k$.} 

With higher attitude maneuverability, AEOS would typically access longer VTW for each target compared to CEOS. Therefore, it is necessary to determine the observation start time for each target. \added{A continuous decision variable $TP_{ik}$ within $[0,1]$ is then introduced to determine the specific satellite observation time for target $i$ on orbit $k$.} \red{}For example, $TP_{ik}=0$ means that the observation start time for target $i$ is $VTS_{ik}$ and when $TP_{ik}$ equals to 1/2, the corresponding observation starts from the middle time of $[VTS_{ik},VTE_{ik}-ot_i]$. The relationship between $TP_{ik}$ and OW is shown as follows.

\begin{gather}
OTS_{ik} = TP_{ik}\cdot(VTE_{ik}-ot_{i}-VTS_{ik}) + VTS_{ik}  \label{EquationOTS} \\	
OTE_{ik} = OTS_{ik} + ot_{i}	\label{EquationOTE}
\end{gather}
        
Set $b_{ik} = 1$ when target $i$ is visible on orbit $k$, and $b_{ik} = 0$ otherwise.
Time intervals $[OTS_{ik},OTE_{ik}]$ and $[VTS_{ik},VTE_{ik}]$ respectively denote the OW and VTW of target $i$ on orbit $k$.  
When the satellite accomplishes an observation mission, a succession of attitude transformation process is required to observe the next target.
Notice the required transition time \added{$Trans(i,OTS_{ik},j,OTS_{jk},k)$} from target $i$ to $j$ is also related to $OTS_{ik}$ and $OTS_{jk}$. We denote the transition time as \added{$Trans(i,j,k)$} for simplicity afterwards.
\red{}Besides, in practical satellite application, a certain period to stabilize the satellite attitude is inevitable. Following~\cite{liu2017adaptive}, the attitude stabilization time is considered as follows.
\begin{equation}
\begin{aligned}
\added{Trans(i,j,k)} = max(|\theta _{ik}^{Pitch} - \theta _{jk}^{Pitch}|/s_k^{Pitch},\\
|\theta _{ik}^{Roll} - \theta _{jk}^{Roll}|/s_k^{Roll}){\rm{ + }}\left\{ {\begin{array}{*{20}{c}}
	5&{}&{\Delta {\rm{g}} \le 15}\\
	{10}&{}&{15 < \Delta {\rm{g}} \le 40}\\
	{15}&{}&{40 < \Delta {\rm{g}} \le 60}
	\end{array}} \right. \label{EquationSt}
\end{aligned}
\end{equation}
\red{}where $\Delta {\rm{g}}$ indicates the total angle change between two adjacent \blue{targets}, $s_{k}^{Pitch}$ and $s_{k}^{Roll}$ represent the attitude maneuvering angle velocity of pitch and roll axes, respectively. During the observation of target $i$, the observation angles of AEOS are denoted as $\theta_{ik}^{Pitch}$ and $\theta_{ik}^{Roll}$. 
The attitude transformation process of AEOS from target $i$ to $j$ also consumes energy, which is described as
\begin{equation}
se_{ij}^{k} = (|\theta_{ik}^{Pitch} - \theta_{jk}^{Pitch}| + |\theta_{ik}^{Roll} - \theta_{jk}^{Roll}|) \cdot e_{k}^{\prime} \label{EquationSe}
\end{equation}

With the parameters defined in Equations~\eqref{EquationOTS}--\eqref{EquationSe}, 
the mathematical model of multi-AEOS scheduling problem is constructed as
\begin{align}
\text{max}\quad& \sum\limits_{i\in{A}} \sum\limits_{k\in{O}} \omega_{i}\cdot x_{ik} \label{ObjFunc}	
\intertext{subject to} & \sum\limits_{k\in{O}} x_{ik} \leq 1 & \forall{i}\in{A} \label{Cons0}  \\
& x_{ik} \leq b_{ik} & \forall{i}\in{A},{k}\in{O} \label{Cons1} \\
& \sum\limits_{i\in{A}} x_{ik} \cdot ot_{i} \cdot m_{k} \leq M_{k} & \forall{k}\in{O} \label{Cons3}
\end{align}
\begin{equation}
\begin{aligned}
    & \sum\limits_{i\in{A}} x_{ik}(se_{ij}^{k} + ot_{i} \cdot e_{k}^{\prime}) \leq E_{k} \\
    & \quad \added{j\text{ is the successor target of } i \text{ on orbit } k,} & \forall{k}\in{O} \label{Cons4}
\end{aligned}
\end{equation}
\begin{equation}
\begin{aligned}
	\{x_{ik} + & x_{jk} \leq 1 |\text{\added{if }}\added{i \text{ is the precursor target}} \\
	&  \added{\text{of }j\text{ on orbit } k}\text{ \added{and} } \\
	& \added{g(TP_{ik},i,TP_{jk},j,k) > 0}\}  \quad \forall{i,j}\in{A},{k}\in{O} \label{Cons2}
\end{aligned}
\end{equation}
\begin{equation}
    x_{ik} \in \{ 0,1\} \qquad \qquad \qquad \forall{i}\in{A},{k}\in{O} 	\label{Cons5}
\end{equation}
 
The objective function~\eqref{ObjFunc} aims to maximize the entire observation profit. 
    Constraints~\eqref{Cons0} and~\eqref{Cons1} represent that each target cannot be observed more than once and should be arranged on available orbits. 
    Memory constraints~\eqref{Cons3} ensure that the memory consumption for observation missions cannot exceed the memory capacity on each orbit. 
    Constraints~\eqref{Cons4} indicate that the sum of energy consumption from satellite maneuvering and imaging should be less than or equal to the energy capacity \added{on} each orbit.
    \added{The attitude} transformation constraints are described in~\eqref{Cons2}, where $i$ is the precursor target of $j$ on orbit $k$. 
    \added{Formula $g(TP_{ik},i,TP_{jk},j,k) > 0$ indicates that the observation ending time of target $i$ plus the attitude transformation time from target $i$ to $j$ is greater than the observation start time of target $j$.}\added{ In line with Eqs.~\eqref{EquationOTS} and~\eqref{EquationOTE},  $g(TP_{ik},i,TP_{jk},j,k)$ is formulated as}
\begin{equation}
\footnotesize
    \begin{aligned}
       & g(TP_{ik},i,TP_{jk},j,k)  = OTE_{ik} + Trans(i,j,k) - OTS_{jk} \\
	& = TP_{ik}\cdot(VTE_{ik}-ot_{i}-VTS_{ik}) + VTS_{ik} + ot_{i} + Trans(i,j,k)   \\
	&   - TP_{jk}\cdot(VTE_{jk}-ot_{j}-VTS_{jk}) - VTS_{jk} \label{Cons-g}
	\end{aligned}
\end{equation}   
\added{where $TP_{ik}$ and $TP_{jk}$ are  independent variables.} If the conditions hold, targets $i$ and $j$ cannot be observed on the same orbit $k$.

\subsection{AEOS scheduling under uncertainty} \label{subsec:reform}

\red{}We address modeling the cloud coverage uncertainty for AEOS scheduling in this subsection, and several necessary assumptions are considered as follows.

\begin{enumerate}
	\item Cloud coverage for targets is simplified to two conditions: complete cloud occlusion and no cloud occlusion. 
	\item For each VTW, cloud occlusion or not is a random event with certain probability. 
	\item The probability of cloud coverage during the VTW is supposed to be the same.
\end{enumerate}

Binary stochastic \red{}parameters $\lambda_{ik}$ are defined to depict whether there is cloud coverage or not. 
The probability that target $i$ can be successfully observed from orbit $k$ is set as $p_{ik}$\red{}, which is randomly generated for each target.
\red{}For each $\lambda_{ik}$, it will be randomly set as 1 with the probability of $p_{ik}$, and 0 otherwise.
Subsequently, we extend the previous deterministic model to a CCP model, in which $1 - \alpha$ represents the predefined confidence interval~\cite{liu2009some}. 
The objective function~\eqref{ObjFunc} is modified and chance constraints are added:
\begin{align}
\text{max}\quad				 & f  \label{NewObjFunc}  
\intertext{subject to}\quad  & P\left\{ \sum\limits_{i\in{A}} \sum\limits_{k\in{O}} \omega_{i} \cdot \lambda_{ik} \cdot x_{ik} \geq f \right\} \geq 1 - \alpha   \label{Cons6}
\end{align}

The new objective function~\eqref{NewObjFunc} is to maximize the entire observation profit under predefined confidence interval $1 - \alpha$. 
The chance \red{}constraint~\eqref{Cons6} restricts the value of $f$. Then the CCP model for multi-AEOS scheduling with cloud coverage uncertainty is formulated as: max $f$, subject to constraints~\eqref{Cons0}--\eqref{Cons5} and~\eqref{Cons6}.

This developed model of multiple AEOSs scheduling problem under cloud coverage uncertainty has several main characteristics and solving difficulties.
First, through the adjustment of the pitch axis, AEOSs could access longer VTW for each target than CEOSs\red{}, which brings new challenges to determine the observation start time. 
\red{}Second, the AEOSs scheduling problem under cloud coverage uncertainty will be more difficult to solve owing to the unfixed mission execution sequence~\cite{WangDemeulemeester-4}. Third, 
as the number of observation targets increases, the scheduling complexity significantly goes up as well, calling for better scheduling heuristics. The AEOSs scheduling problem under cloud coverage uncertainty is therefore more difficult than that for non-agile satellites, \red{}due to the stronger maneuverability and more complicated calculations of attitude transformation time and energy consumption.

\red{}Besides, as far as we can see, there exists no algorithm designed specifically for our proposed model.
	Although there are well-designed algorithms for classic AEOSs scheduling problem (e.g.,~\cite{liu2017adaptive,peng2019agile}), introducing cloud coverage uncertainty could further increase the complexity for scheduling results evaluation, and consequently require more effective solution update process. 
	Moreover, part of these existing algorithms do not consider energy constraints, which could also increase the problem solving difficulty.
	To better address the uncertain AEOSs scheduling model, designing a novel heuristic being capable of efficient solution update and handling memory and energy constraints is required.

\section{Solution method}
\label{sec:method}

In this section, \red{}the sample approximation method~\cite{luedtke2008sample} is initially introduced to calculate
confidence profit considering cloud coverage uncertainty. The optimization subproblem
of selecting the start time of \red{}OW is then defined and solved by \red{}the SQP method. 
Finally, \red{}an ISA based heuristic integrating a fast \red{}insertion algorithm
is proposed to maximize the observation profit under predefined confidence interval.

\subsection{Sample approximation method}
It is difficult to calculate the probability in the chance constraint~\eqref{Cons6}. 
Therefore, a sample approximation method is adopted to calculate observation profit approximately.
Monte Carlo simulation is a numerical calculation method guided by probability and statistics theory~\cite{metropolis1949monte}.
This statistical method generates a set of scenarios to describe situations with different cloud coverage. The scenarios can be depicted as $W = \{ w_{0},w_{1},...,w_{n}\}$, in which any $w_{l}\in W$ corresponds to a group of $\lambda_{ik}$, and $|W|$ \red{}denotes the sample size. 

Denote $1-\epsilon$ as the confidence level of solution, which reflects the proportion of scenarios that satisfy the feasible solution of sample approximation problem. In line with~\cite{ruszczynski2002probabilistic}, the solution which is infeasible for at most $|W|\cdot \epsilon$ scenarios can be obtained. 
Introduce $y_{l}$ as binary variables: $y_{l} = 0$ if the solution is feasible for $w_{l}$ and $y_{l} = 1$ otherwise. The sample approximation constraints are described as

\begin{equation}
\sum\limits_{i\in{A}} \sum\limits_{k\in{O}} \omega_{i} \cdot \lambda_{ik}^{l} \cdot x_{ik} \geq -y_{l} \cdot M + f \qquad  \forall w_{l}\in W  \label{SACons1}	\\
\end{equation}
\begin{equation}
\sum\limits_{w_{l}\in W} y_l \leq |W|\cdot \epsilon     \label{SACons2}
\end{equation}

In constraints~\eqref{SACons1}, $\lambda_{ik}^{l}$ denotes the value of parameter $\lambda_{ik}$ under scenario $w_{l}$ and $M$ is assumed to be a large number. If $y_{l} = 0$, constraints~\eqref{SACons1} indicate that the total profit of observation must be larger than or equal to $f$. \red{}When the obtained solution is not feasible (i.e., its observation profit is less than $f$), the binary variable $y_l$ will be taken as 1 for each $w_l \in W$, which ensures constraints~\eqref{SACons1} hold for the scheduling solution. 
Constraint~\eqref{SACons2} imposes that the number of scenarios in which the solution is infeasible should be at most $|W|\cdot \epsilon$. \red{}Therefore, to obtain a higher observation profit, the solution could be infeasible in $|W|\cdot \epsilon$ scenarios, while its observation profit in the remaining scenarios should be no less than $f$. 
The entire observation profit $f$ under predefined confidence interval, \red{}which is the optimization function in the CCP model, now can be \red{}determined.

A lower bound of the sample size $|W|$ has been proposed in~\cite{luedtke2008sample}, and is expressed as
\begin{align}
|W| \geq \frac{1}{2( \epsilon-\alpha)^{2}}  log(\frac{1}{\theta}) + \frac{\beta}{2(\epsilon-\alpha)^{2}}  log(U)           \label{SampleSize}
\end{align}
where $1- \epsilon$ is greater than $1- \alpha$. 
$1-\theta$ is the probability when the feasible solution of the sample approximation problem simultaneously meets the original CCP model. $U$ is set as $|X|^{1/\beta}$, where  $|X|$ is the norm of the decision variables vector and $\beta$ is the number of decision variables.

\subsection{Optimization subproblem}  
\label{subsec:opti-sub}
\red{}In order to schedule more observation missions and obtain higher observation profit, the range of available time intervals between consecutive observation missions should be appropriately determined. 
To achieve this, the concept of time slack from~\cite{liu2017adaptive} is introduced to describe the time intervals, and the observation start time is consequently determined by the SQP method.

\subsubsection{Definition of time slacks}
Time slacks denote the time interval between two adjacent \blue{targets} on the same orbit, \red{}so the subscript $k$ is omitted here.

\begin{figure}[htb]
	\begin{center}
		\includegraphics[width=0.46\textwidth]{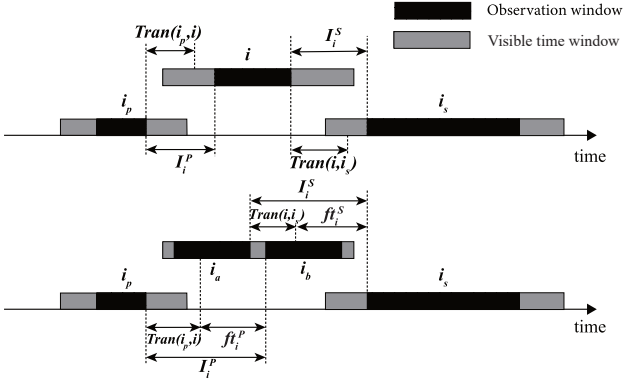}
		\caption{Calculation diagram of \red{}the time slack.}\label{fig:TimeSlack}
	\end{center}
\end{figure}

As shown in Fig.~\ref{fig:TimeSlack}, we intend to insert \blue{target} $i$ between its precursor \blue{target} $i_{p}$ and successor \blue{target} $i_{s}$ for better scheduling results. 
\red{}The $i_a$ and $i_b$ are the example insertion \blue{targets} in two different situations, corresponding to the time slacks for the successor and precursor \blue{targets} respectively.
\red{}Two time slacks are denoted as $ft_{i}^{S}$ and $ft_{i}^{P}$. 
\red{}During the insertion process, where the time slack is utilized, the observation start times of successor and precursor \blue{targets} of $i$ are known.
The value of $Trans(i,i_{s})$ \red{}thus varies with $OTS_i$,
and $ft_{i}^{S}$ denotes the time slack between \blue{targets} $i$ and $i_{s}$, which is defined as 
\begin{equation}
ft_{i}^{S} = max\left(I_{i}^{S}-Trans(i,i_{s})\right) \label{eq:ft_{i}^{S}}
\end{equation}
where $I_{i}^{S}$ is the time interval from  $i$ to $i_{s}$ and
\begin{equation}
I_{i}^{S} = OTS_{i}^{S}-OTE_{i}	\label{eq:I_{i}^{S}}
\end{equation}
where $OTS_{i}^{S}$ represents the observation start time of \blue{target} $i_{s}$. \red{}The independent variable of Equation~\eqref{eq:ft_{i}^{S}} is $OTS_i$, which varies in the range of $[ VTS_{i},VTE_{i} - ot_{i}]$.

Substituting~\eqref{EquationOTS} and~\eqref{EquationOTE} into~\eqref{eq:I_{i}^{S}}, $I_{i}^{S}$ is expressed as
\begin{equation}
I_{i}^{S} = OTS_{i}^{S}-\left[TP_{i} \cdot (VTE_{i}-ot_{i}-VTS_{i}) + VTS_{i} + ot_{i}\right]  \label{eq:I_{i}^{S}_1}
\end{equation}

If \blue{target} $i$ is the last observation \blue{target} on the orbit, then 
\begin{equation}
\begin{aligned}
ft_{i}^{S} &= VTE_{i} - ot_{i} -OTS_{i} \\
&= (1-TP_{i}) \cdot (VTE_{i}-ot_{i}-VTS_{i})
\end{aligned} 
\end{equation}  

Therefore $ft_{i}^{S}$ is the function of decision variable $TP_{i}$, which suggests that the time slack is determined by the \red{}start time of OW. 

The time slack between \blue{target} $i$ and its precursor \blue{target} $i_{p}$, denoted as $ft_{i}^{P}$, can be calculated similarly. If \blue{target} $i$ is not the first \blue{target} on \red{}current orbit, then
\begin{equation}
ft_{i}^{P} = max\left(I_{i}^{P}-Trans(i_{p},i)\right)
\end{equation}
where $I_{i}^{P}$ is calculated as
\begin{equation}
\begin{aligned}
I_{i}^{P} = &OTS_{i} - OTE_{i}^{P} = [TP_{i} \cdot (VTE_{i}-ot_{i}-VTS_{i}) \\
&+ VTS_{i}] - OTE_{i}^{P}
\end{aligned}
\end{equation}
where $OTE_{i}^{P}$ denotes the observation \added{ending} time of \blue{target} $i_{p}$.

If \blue{target} $i$ is the first \blue{target} on the orbit, then 
\begin{equation}
ft_{i}^{P} = OTS_{i} - VTS_{i} = TP_{i} \cdot (VTE_{i}-ot_{i}-VTS_{i})
\end{equation}

\subsubsection{Determining the \red{}start time of OW}
To determine the \red{}start time of OW, \red{}namely the proper insertion position of the \blue{target}, the optimization subproblem is constructed as follows.
\begin{align} 
\text{\red{}min}\quad& \left( TP_{i} - 1/2\right)^{2}		\label{SubObjFunc}	 \\	
\intertext{subject to} & I_{i}^{S}-Trans(i,i_{s}) \geq 0 & \label{constr:trans_s}\\ 
& I_{i}^{P}-Trans(i_{p},i) \geq 0 & \label{constr:trans_p}\\ 
& 0 \leq TP_{i} \leq 1 & 
\end{align}

The objective function~\eqref{SubObjFunc} makes $TP_{i}$ as close to 1/2 as possible. 
Constraints~\eqref{constr:trans_s} and~\eqref{constr:trans_p} represent the attitude transformation time constraints when \blue{target} $i$ has the precursor and successor \blue{targets}.
\red{}During the process of determining the observation start time of \blue{target} $i$, which is critical to the insertion strategy, the OWs for the precursor and successor \blue{targets} are known. According to the definitions of time interval and \blue{attitude transformation} time, the unique independent parameter of the optimization subproblem is $TP_i$. 
The optimization model has the following advantages: (1) When the value of $TP_{i}$ is around 1/2, the satellite would execute observation missions above the target $i$, which 
ensures higher observation resolution. 
(2) The Euclidean distance~\cite{danielsson1980euclidean} between $TP_{i}$ and $1/2$ is adopted to ensure that the objective function is convex and thus efficient algorithms can be applied.

\red{}The optimization subproblem is established on the purpose of observing more targets and obtaining higher observation profit. Meanwhile, it is desirable to obtain high-resolution images,  although we do not consider the observation imaging quality as our objective function. 
The \red{}suitable insertion position can be determined by solving the above quadratic optimization model and the actual observation time window is obtained simultaneously.
The SQP method, which performs well in solving nonlinearly constrained optimization problems~\cite{boggs1995sequential}, is employed to solve this optimization subproblem.

\subsection{ISA based heuristic}
The design of heuristic rules is of great importance for the effectiveness and efficiency of the proposed algorithm. With effective heuristic rules, we start with an initial solution and gradually approach \red{}the solution via ISA combining with a fast insertion algorithm.

\subsubsection{Selection rules of \blue{targets} and resources}
\label{subsec:rules}
Before operating the fast insertion algorithm, 
the observation \blue{target} and corresponding resource should be selected. 
The urgency for each target $Need_{i}$ is defined as follows.
\begin{equation}
Need_{i} =  \frac{\omega_{i}}{\omega_{max}} + \left( 1 - \frac{\sum\limits_{k\in{O}}  p_{ik}}{N_{i}}\right) 
\end{equation}
where $\omega_{max}$ represents \red{}the largest observation profit of a single target among all the observation targets and $N_{i}$ denotes the number of VTWs of target $i$.

As seen in the definition of $Need_i$, the first item stands for the potential normalized observation profit, while the second item indicates the successful observation probability of target $i$. Therefore the higher value of $Need_{i}$ corresponds to higher observation priority for target $i$.   

The corresponding observation resource for the target is determined on the basis of \red{}the resource selection rule.  Considering that the targets could be observed on the same orbit $k$, the OWs of different targets may overlap and the energy/memory consumption may exceed the orbit capacity. For each target $i$ on orbit $k$, the degree of resource conflict denoted as $CF_{ik}$ is then defined as
\begin{equation}
CF_{ik} = (1-p_{ik}) \cdot \left(\frac{CFS_{ik}}{|VTW_{ik}|}+\sum\limits_{r=1}^{R} \frac{Cost_{irk}}{Cap_{rk}}\right)  \label{def:Conf}
\end{equation}
where $CFS_{ik}$ denotes the length of overlapping intervals between target $i$ and other targets on orbit $k$. \red{}The duration of VTW for target $i$ on orbit $k$ is denoted as $|VTW_{ik}|$ and $R$ is the number of \red{}resource constraint ($R=2$ in this work since the energy and memory constraints are considered). 
The consumption of resource $r$ for target $i$ on orbit $k$ is represented as $Cost_{irk}$, and the remaining amount of resource $r$ is denoted as $Cap_{rk}$. 
\red{}In Equation~\eqref{def:Conf}, parameter $p_{ik}$ indicates the probability that target $i$ will be successfully observed on orbit $k$. 
	A larger $p_{ik}$ means that target $i$ is more likely to observe successfully. 
	Then $CF_{ik}$ will be lower, which means that the corresponding resource will be allocated to the target earlier. 
	The second part $\dfrac{{CF{S_{ik}}}}{{|VT{W_{ik}}|}}$ represents the proportion of the VTW of target $i$ that overlaps with other VTWs on orbit $k$.
	The third part, $\sum\limits_{r = 1}^R {\dfrac{{{\mathop{\rm Cos}\nolimits} {t_{irk}}}}{{Ca{p_{rk}}}}}$, represents the proportion of the resource consumed by observing the target in the remaining resource. The larger values of these two parts, the later the observation of corresponding targets will be considered and scheduled. When $Cost_{irk}$ is larger than $Cap_{rk}$, the target $i$ will not be arranged to observe with the current satellite orbit. Simultaneously, this criterion also guarantees the energy constraints and memory constraints.

\subsubsection{Fast insertion algorithm}
\begin{figure}[htbp]
	\begin{center}
		\includegraphics[width=0.50\textwidth]{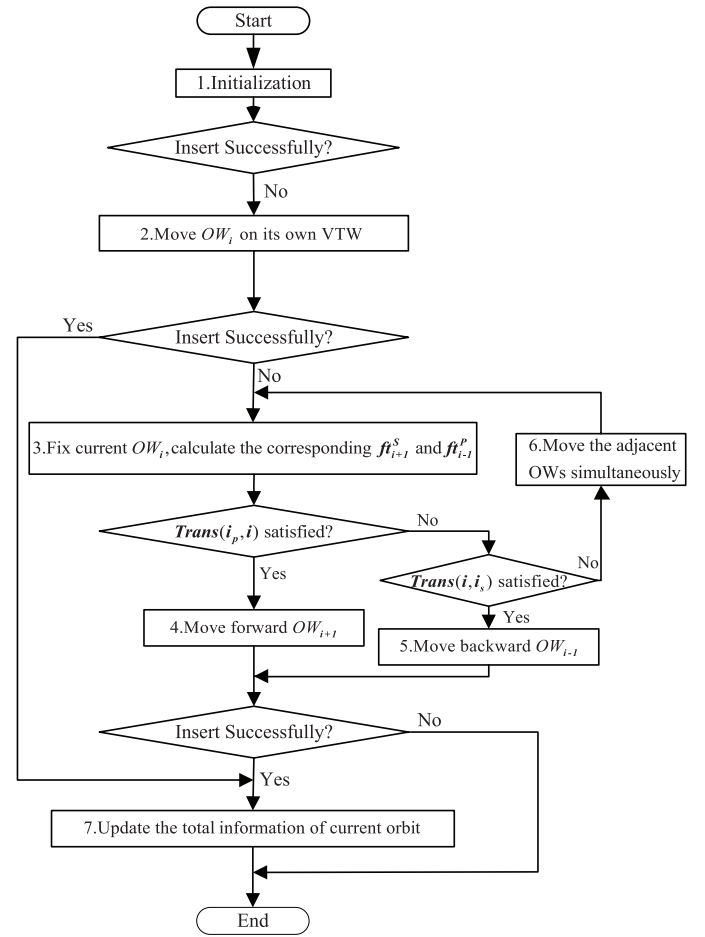}
		\caption{The flow chart of the fast insertion algorithm.}\label{figInsert}
	\end{center}
\end{figure}

Suppose that \blue{target} $i$ is to be executed on orbit $k$, and there are already two scheduled \blue{targets}, the successor \blue{target} $i+1$ and the precursor \blue{target} $i-1$, on the same orbit. The fast insertion algorithm is then invoked to insert \blue{target} $i$ between the two scheduled \blue{targets}. 
The main procedure of the fast insertion algorithm is summarized in Fig.~\ref{figInsert}, and the detailed procedure is described as follows.

\textbf{Step 1}: During the initialization process, $TP_{ik}$ is set as 1/2, 
and the parameters $I_{i}^{P}$, $I_{i}^{S}$, $Trans(i_{p},i)$ and $Trans(i,i_{s})$ are calculated.
Then we check the constraints~\eqref{Cons3}--\eqref{Cons5} to determine $x_{ik}$. 

\textbf{Step 2}: If the initial OW for \blue{target} $i$ (denoted as $OW_i$) does not satisfy the transition time constraint, shuffle $OW_i$ within the VTW of \blue{target} $i$ on orbit $k$ in order to search a proper position that meets the constraint. Through solving the optimization subproblem defined in Section~\ref{subsec:opti-sub}, the value of $TP_{ik}$ corresponding to the position of $OW_i$ is updated if constraints ~\eqref{constr:trans_s} and~\eqref{constr:trans_p} are simultaneously satisfied\red{}, then go to Step 7. Otherwise, go to Step 3.

\textbf{Step 3}: \red{}Fix $OW_i$ at the initial position, namely the value of $TP_{ik}$ equals to 1/2, and calculate the subsequent time slack of $OW_{i+1}$ and the precursor time slack of $OW_{i-1}$, which are denoted as $ft_{i+1}^{S}$ and $ft_{i-1}^{P}$, respectively. If constraint~\eqref{constr:trans_p} holds, move forward $OW_{i+1}$ and go to Step 4. If constraint~\eqref{constr:trans_s} holds, move backward $OW_{i-1}$ and go to Step 5. If none of them holds, go to Step 6. 

\textbf{Step 4}: 
If condition $ft_{i+1}^{S} \geq I_{i}^{S} - Trans(i,i+1)$ holds, move forward the successor observation time window $OW_{i+1}$. If $OW_{i}$ can be inserted successfully, go to Step 7. Otherwise, current insertion fails and the algorithm ends. 

\textbf{Step 5}: 
If condition $ft_{i-1}^{P} \geq I_{i}^{P} - Trans(i-1,i)$ holds, move backward the precursor observation time window $OW_{i-1}$. If $OW_{i}$ can not be inserted, end the current insertion procedure. Otherwise, go to Step 7.

\textbf{Step 6}: 
If conditions $ft_{i+1}^{S} \geq I_{i}^{S} - Trans(i,i+1)$ and $ft_{i-1}^{P} \geq I_{i}^{P} - Trans(i-1,i)$ hold simultaneously, insert $OW_{i}$ by moving the two adjacent OWs at the same time (Repeat Step 4 and Step 5). If the above conditions hold, go to Step 7. Otherwise, end current insertion process.      

\textbf{Step 7}: Once $OW_{i}$ has been inserted successfully, update the information of resource consumption on orbit $k$.

\subsubsection{Structure of the \red{}ISA based heuristic}

\begin{figure}[htbp]
	\begin{center}
		\includegraphics[width=0.4\textwidth]{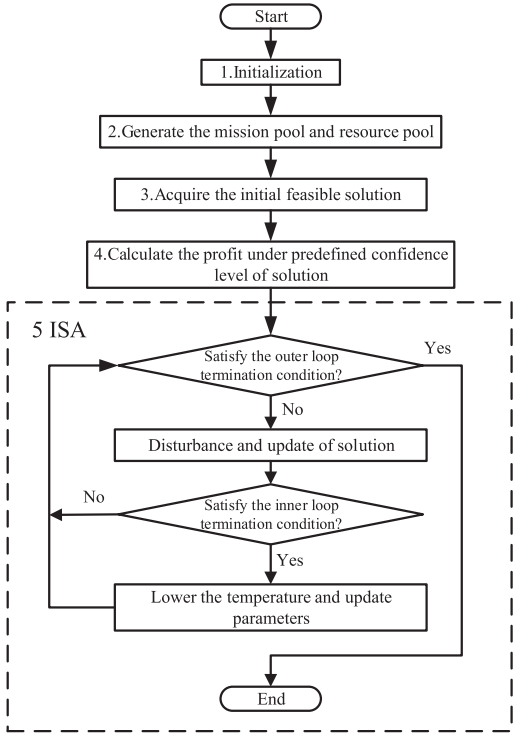}
		\caption{The structure of ISA based heuristic.}\label{figSAheuristic}
	\end{center}
\end{figure}

\red{}Metaheuristic algorithms have been extensively employed in various practical engineering problems~\cite{wang2018multi,zhang2019large,xiang2019comprehensive}. Particularly, the \red{}simulated annealing algorithm proposed decades ago~\cite{kirkpatrick1983optimization, kirkpatrick1984optimization}, is a stochastic optimization method. Possessing the ability to acquire near-optimal solutions within an acceptable time, simulated annealing has been widely applied in the combinatorial optimization problems.
Combining the fast insertion strategy, we develop \red{}an ISA based heuristic for multi-AEOS scheduling under cloud coverage uncertainty. The primary structure of \red{}ISA based heuristic is shown in Fig.~\ref{figSAheuristic}, where each step is  described as follows.

\textbf{Step 1}: Initialize all input parameters, such as satellite orbital elements, the information of targets and the VTWs. \red{}Besides, obtain an initial scheduling result using the proposed selection rules of \blue{targets} and resources in Section~\ref{subsec:rules}.

\textbf{Step 2}: Calculate $Need_{i}$ and $CF_{ik}$ for each target $i$ on available orbit $k$. By sorting these two factors in order, generate the initial mission pool and resource pool.

\textbf{Step 3}: Generate initial feasible solution. Select targets to be observed from the mission pool and corresponding available resources in order. The resource with lower degree of conflict has higher priority. If the target is successfully scheduled, delete related target and resource in the pools. 

\textbf{Step 4}: \red{}Calculate confidence observation profit via the sample approximation method \red{}for the initial solution.

\textbf{Step 5}: The dotted box contains the general procedure of ISA, which is detailed in Algorithm~\ref{improvedSA}. The input parameters of ISA are given initially, where higher $\gamma$ means larger disturbance of current solution, minor $\zeta_{m}$ indicates lower increasing speed of $nF$, which denotes the unaccepted numbers of new solution.
\red{}$nF^{T}_{m}$ is the upper limit of the iterations number in inner loop, while $nF_{m}$ and $nIter_{m}$ are set as the maximum numbers of iterations in outer loop. 

During each iteration, a certain number of targets would be deleted and a new solution is generated with randomly selecting targets from the mission pool and invoking the fast insertion algorithms. Using the popular Metropolis method~\cite{kirkpatrick1983optimization}, the acceptance probability $P_{a}$ is designed as 
\begin{equation}
P_{a} = \left\{ 
\begin{array}{rcl}
1						&{\Delta f \geq 0}\\
e^{\Delta f / T}		&{\Delta f < 0}
\end{array}
\right.   \label{eq:Pa}
\end{equation}  
where $\Delta f = f^{new} - f$. With the help of Metropolis strategy, ISA can jump out of local optima to some extent. Subsequently, a updating strategy for $nF^{T}$ is proposed:

\begin{equation}
nF^{T} = \left\{ 
\begin{array}{rcl}
0					&	&{\zeta \geq \zeta_{m}}\\
nF^{T} + 1	&	&{\zeta < \zeta_{m}}
\end{array}
\right.   \label{eq:nF} 
\end{equation}
where \red{}$T$ is a mark of the current temperature, which does not participate in calculations, and $\zeta$ indicates the improving ratio of \red{}the new solution compared to the old one\red{}, calculated as
\begin{equation}
\zeta = (f^{new} - f) / f
\end{equation}

When the inner loop ends, current temperature would decrease due to $\alpha_{T} < 1$. Conversely, \red{}a longer length of Markov chain is updated by $\alpha_{L} > 1$
. Finally, we output the best scheduling solution $S_{best}$ and corresponding observation profit $f_{best}$. \red{}The complexity of the proposed algorithm has been evaluated in Table~\ref{tab:BigONotation}, indicating the total time computational complexity is O($n^2$), where $n$ denotes the number of targets.

\renewcommand{\algorithmicrequire}{ \textbf{Input:}} 
\renewcommand{\algorithmicensure}{ \textbf{Output:}} 
\begin{algorithm}[ht]
	\caption{Procedure of ISA}
	\label{improvedSA}
	\footnotesize
	\begin{algorithmic}[1]
		\REQUIRE ~~\\
		Information of initial profit $f_{0}$ and number of visible targets $vTar$, initial scheduling scheme $S_{0}$ and the number of \blue{targets} arranged $aTar^{0}$, parameters of ISA: $\zeta_{m}$, $\gamma$, $\alpha_{T}$, $\alpha_{L}$, $T_{0}$, $nF_{m}$, $nF^{T}_{m}$ and $nIter_{m}$
		\ENSURE  ~~\\
		The best total profit $f_{best}$ and corresponding scheduling scheme $S_{best}$ 
		\STATE Initialization: $aTar^{T} \leftarrow aTar^{0}$, $T \leftarrow T_{0}$, $L^{T} \leftarrow vTar / 2$, $f \leftarrow f_{0}$, $f_{best} \leftarrow f_{0}$, $S \leftarrow S_{0}$, $S_{best} \leftarrow S_{0}$; \
		\WHILE{$nF < nF_{m}$ or $nIter < nIter_{m}$}
		\label{L3}
		\WHILE{$nF^{T} < nF^{T}_{m}$ or $nIter^{T} < L^{T}$}
		\STATE Delete $\gamma \times aTar^{T}$ \blue{targets} from current solution randomly and generate a new scheduling scheme $S^{new}$ including $f^{new}$, $aTar^{new}$; \
		\STATE Generate a uniform random number $u$ in [0,1] 
		\IF{$u < P_{a}$}
		\STATE $f \leftarrow f^{new}$, $S \leftarrow S^{new}$, $aTar^{T} \leftarrow aTar^{new}$
		\STATE update $nF^{T}$ according to~\eqref{eq:nF} 
		\IF{$f > f_{best}$}
		\STATE $f_{best} \leftarrow f$, $S_{best} \leftarrow S$
		\ENDIF
		\ELSE 
		\STATE $nF^{T} \leftarrow nF^{T} + 1$
		\ENDIF
		\STATE $nIter^{T} \leftarrow nIter^{T} + 1$
		\ENDWHILE
		\STATE $nF \leftarrow nF + nF^{T}$
		\STATE $nIter \leftarrow nIter + nIter^{T}$
		\STATE $T \leftarrow T \times \alpha_{T}$, $L^{T} \leftarrow L^{T} \times \alpha_{L}$
		\STATE $nF^{T} = 0$, $nIter^{T} = 0$ 
		\ENDWHILE
		\label{L4}
		\STATE Output $f_{best}$ and $S_{best}$\
	\end{algorithmic}
\end{algorithm}

\begin{table}[htbp]
	\centering
	\caption{The complexity of the proposed algorithm.}
	\footnotesize
	\scalebox{0.9}{
	\begin{tabular}{ll}
		\toprule
		Procedure of Algorithm & Complexity \\
		\midrule
		Initialization & O$(1)$ \\
		Generate the mission pool and resource pool & O$(n)$ \\
		Acquire the initial feasible solution & O$(n^2)$ \\
		Calculate the profit under predefined confidence level of solution & O$(n^2)$ \\
		Disturbance and update of solution & O$(n^2)$ \\
		Lower the temperature and update parameters & O$(1)$ \\
		\bottomrule
	\end{tabular}}%
	\label{tab:BigONotation}%
\end{table}%

\section{Computational experiments}
\label{sec:comput}


\subsection{Data generation}
\label{susec:datagene}

The experiments are conducted using Intel (R) Core (TM) i7-4790K CPU at 4.00 GHz and 12.0 GB of RAM on Windows 10 64-bits OS. The \red{}ISA based heuristic is coded in Matlab
and the scheduling scenarios are generated as follows.
Without benchmark dataset for \red{}uncertain AEOSs scheduling problems, we design several instances in line with~\cite{liu2017adaptive,HE201812,han2018scheduling}. The targets are generated according to a random distribution over the world and several specific interest areas. The total number of observation targets is 500, 650, 800 or 950, with a worldwide distribution of 500 targets and additionally several areas with 150 targets each. The profit $\omega_{i}$ of each target $i\in A$ is uniformly distributed from 1 to 10. 
The worldwide distribution targets locate in the range of  latitude between 60$^{\circ}$ S-60$^{\circ}$ N and longitude between 180$^{\circ}$ W-180$^{\circ}$ E. Several interest regions are predefined mainly in China (3$^{\circ}$ N-53$^{\circ}$ N and 74$^{\circ}$ E-133$^{\circ}$ E), Australia (43$^{\circ}$ S-10$^{\circ}$ S and 112$^{\circ}$ E-154$^{\circ}$ E), and America (24$^{\circ}$ N-49$^{\circ}$ N and 73$^{\circ}$ W-125$^{\circ}$ W), among which one, two or three regions are selected for different scenarios. The constraint that the solar altitude angle for each observation is not less than 0$^{\circ}$ is also taken into consideration in generating VTWs.

The mission horizon is set as 24 hours, starting from 2017/01/01 00:00:00. The orbital parameters of satellites in the scenarios are shown in Table~\ref{tab:SatParameters}. The first column $ID$ denotes the name of satellite, and the other columns indicate the length of semi-major axis ($a$), eccentricity ($e$), inclination ($i$), right ascension of the ascending node ($\Omega$), argument of perigee ($\omega$) and mean anomaly ($M$), respectively. The satellites are designed with the largest pitch degree of 30$^{\circ}$ and roll degree of 30$^{\circ}$
, where the angular velocity are all set as 3$^{\circ}/s$. The unit-time memory consumption of imaging $m_{k}$ is 100 MB/s, while the energy consumption of unit-time imaging and unit-angle maneuvering are 500 and 1000 $W$, respectively. In line with~\cite{WangDemeulemeester-4}, the parameters of the CCP are fixed as $1-\alpha=0.90, 1-\epsilon=0.99$ and $1-\theta = 0.99$. 
\red{}The value of $|W|$ will be set as the integer lower bound determined by Equation~\eqref{SampleSize}.
The default parameters of ISA are set as $\alpha_{T} = 0.95$, $\alpha_{L} = 1.05$, $\zeta_{m} = 0.05$, $T_{0} = 1000$, $nF^{T}_{m} = 50$ and $nIter_{m} = 2000$.

\begin{table}[htbp]
	\footnotesize
	\centering
	\caption{Orbital parameters of the satellites.}
	\label{tab:SatParameters}
	\scalebox{0.9}{
	\begin{tabular}{crrrrrr}
		\toprule
		$ID$  & \multicolumn{1}{c}{$a$(km)} & \multicolumn{1}{c}{$e$} & \multicolumn{1}{c}{$i$($^{\circ}$)} & \multicolumn{1}{c}{$\Omega$ ($^{\circ}$)} & \multicolumn{1}{c}{$\omega$ ($^{\circ}$)} & \multicolumn{1}{c}{$M$ ($^{\circ}$)} \\
		\midrule
		Sat1  & 6903.673  & 0.001655  & 97.5839  & 97.8446  & 50.5083  & 2.0288  \\
		Sat2  & 6903.730  & 0.001558  & 97.5310  & 95.1761  & 52.2620  & 31.4501  \\
		Sat3  & 6909.065  & 0.000997  & 97.5840  & 93.1999  & 254.4613  & 155.2256  \\
		Sat4  & 6898.602  & 0.001460  & 97.5825  & 92.3563  & 276.7332  & 140.1878  \\
		\bottomrule
	\end{tabular}}%
\end{table}%

\subsection{Computational results}

\subsubsection{Parameters setting}

\begin{figure*}[htbp]
	\begin{center}
		\includegraphics[width=0.75\textwidth]{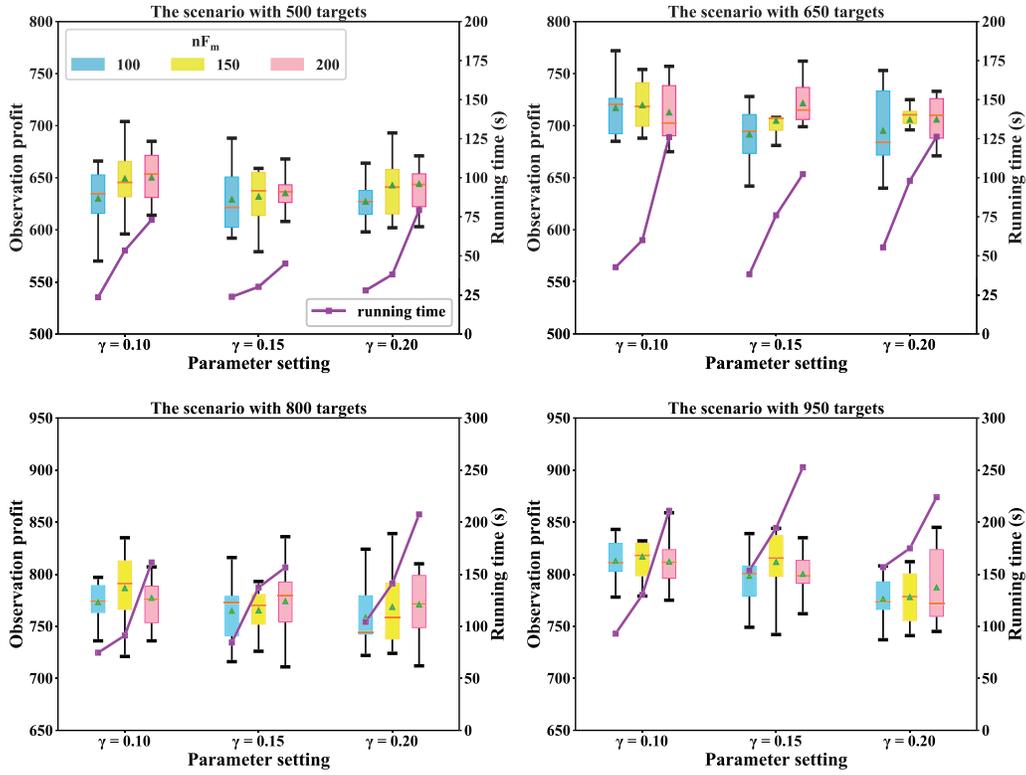}
		\caption{{\bfseries} Simulation results of parameter setting with different number of targets. }\label{figGamma}
	\end{center}
\end{figure*}

In the proposed algorithm, two parameters of the disturbance rate $\gamma$ and the maximum iteration number $nF_{m}$ play important roles in the acquisition of better results. The disturbance rate $\gamma$ for arranged \blue{targets} would affect the scheduling variation range, and $nF_{m}$ controls the termination condition. The tuning experiments for $\gamma$ and $nF_{m}$ have been conducted where the energy and memory capacity \added{on} each orbit are set as $E_{m}$ = 80 kJ and $M_{m}$ = 7500 MB.

\red{}The simulation results for different scenarios (n = 500, 650, 800, 950) are shown in Fig.~\ref{figGamma}.
	The green triangle symbol and the orange horizontal line indicate the average and median of the observation profit of 10 runs, respectively. The box plot shows the distribution of the results and the purple square dots represent the algorithm running time.
	As the number of observation targets increases, the observation profit under each set of parameters improves gradually, while the program running time also has an ascending trend. 
	Compared to the instance with $\gamma = 0.10$ and $nF_{m} = 100$,
	the observation profit of other groups does not have generally evident increase and even slightly decrease,
	while the program running time has improved significantly in all scenarios with different number of targets.
	In order to strike a balance between the performance and efficiency of \red{}the algorithm, the set of parameters $\gamma = 0.10$ and $nF_{m} = 100$ is therefore selected for the following experiments in multiple AEOSs scheduling problem with cloud coverage uncertainty.

\subsubsection{Constraint testing}

\begin{figure*}[htbp]
	\begin{center}
		\includegraphics[width=0.76\textwidth]{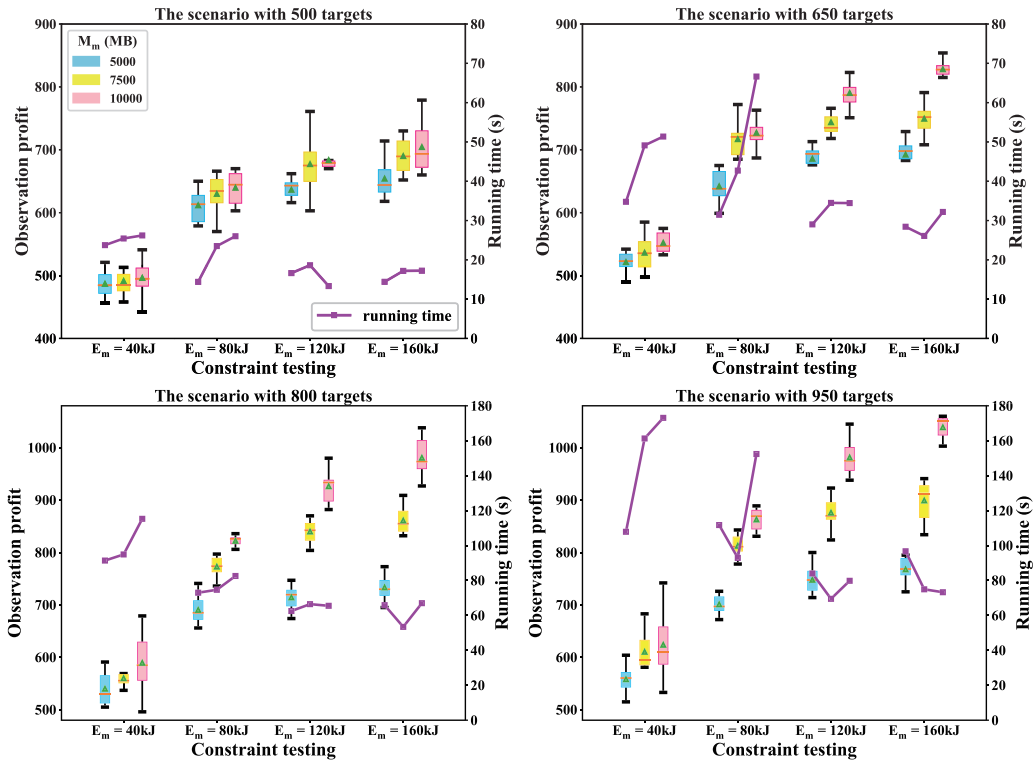}
		\caption{{\bfseries} Simulation results of constraint testing with different number of targets.}\label{figEMAll}
	\end{center}
\end{figure*}

\red{}The energy and memory constraints are crucial for multi-AEOS scheduling with uncertainty problem. We test the constraints with different parameters and report the simulation results in Fig.~\ref{figEMAll}. As shown in the four subplots, the observation profit clearly increases with more number of targets.
	When $E_{m}$ is fixed, the observation profit improves with the increase of $M_{m}$. Similarly, for the instances with the same $M_{m}$, the profit also goes up as $E_{m}$ increases, indicating the influence of the constraints for the multi-AEOS scheduling problem.

\red{}For the scenario with 500 targets whose results are shown in the top left subplot, the most significant profit improvement appears when $E_{m}$ varies from 40 to 80 kJ, and the average increase of mean observation profit in conditions of different memory capacity ($M_{m}$ = 5000, 7500 and 10000 MB) is about 150. When $E_{m}$ is set as 120 kJ or 160 kJ, we observe that the observation profit increases slowly, indicating that the energy capacity with 80 kJ \added{on} each orbit may be large enough for scheduling in this scenario.

\red{}When the energy constraint is not binding anymore for the AEOSs scheduling, the memory capacity would be the most important factor for the scheduling result.
	It can be observed from the two subplots below.
	For the scenarios with $E_{m}$ setting as 120 and 160 kJ, the observation profit dramatically increases with the increase of the memory capacity. This is because that the energy is sufficient for observation and the memory storage is not enough in current situations.
	In conclusion, the energy and memory constraints have a huge impact on the observation profit, and the problem can be effectively solved by the proposed heuristic.

\subsubsection{Comparison experiments with other heuristics}

To verify the effectiveness of the proposed algorithm, one
should compare it with existing algorithms. However, as mentioned in the introduction, no existing algorithms can be readily applied to the multiple AEOSs scheduling problem under cloud coverage uncertainty. \red{}Therefore we introduce the genetic algorithm (GA), the adaptive large neighborhood search (ALNS) algorithm proposed in~\cite{liu2017adaptive}, the bidirectional dynamic programming based iterated local search (BDP-ILS) algorithm presented in~\cite{peng2019agile}, and modify them for comparisons.
The original observation profit in ALNS is redefined as the profit under certain confidence level, and for fair comparison, the maximum iteration number of ALNS is changed to 100 and 500 (denoted as ALNS-100 and ALNS-500, respectively), remaining other parameters the same as in~\cite{liu2017adaptive}. 
\red{}Apart from the adjustment of the uncertain profit, the memory constraint is evaluated during the \blue{target} insertion and the energy consumption has been calculated for each solution in the BDP-ILS algorithm. Moreover, in order to ensure comparable algorithm run time with ISA, the iteration number and the removal ratio of BDP-ILS are determined as 100 and 0.1, respectively.
A similar process with the ISA is applied in the GA, by initializing the scheduling results with the fast insertion algorithm and executing crossover and mutation rules during each iteration. According to preliminary experiments, the parameters of GA are determined as follows.

\begin{itemize}
	\setlength{\itemsep}{0pt}
	\setlength{\parsep}{0pt}
	\setlength{\parskip}{0pt}
	\item Population size: 10
	\item Crossover probability of population individual: 0.5
	\item Crossover and mutation criterion: single point
	\item Iteration: 100
\end{itemize}

\begin{figure}[htbp]
	\begin{center}
		\includegraphics[width=0.48\textwidth]{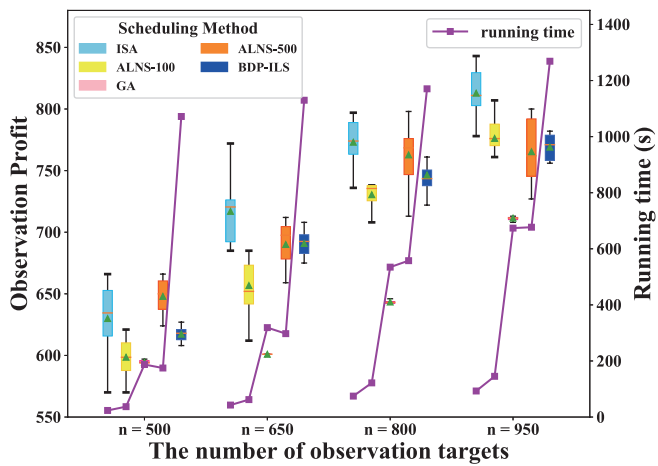}
		\caption{{\bfseries} 
			Comparison experiments results with different algorithms. }\label{figComparative}
	\end{center}
\end{figure}

The results of the comparative experiments are shown in Fig.~\ref{figComparative}, in which the green triangle symbol and the orange horizontal line indicate the average and median values of the observation profit of 10 runs, respectively. The box plot shows the distribution of the results and the purple square dots represent the algorithm running time.
As can be seen from Fig.~\ref{figComparative}, the mean observation profit of each scenario obtained by ISA is higher than \red{}that of ALNS-100, GA and BDP-ILS, while the program running time of ISA is the shortest.
Although GA consumes \blue{}a longer computation time, 
the scheduling profit is worse than the results of ISA and ALNS-100/500. This is because the crossover and mutation strategies contribute little to the generation of high-quality scheduling schemes, resulting in stable solutions. 
\blue{}The ALNS algorithm will exit the loop immediately when all targets are arranged for observation, which is efficient for deterministic AEOS scheduling while not the case for uncertain scheduling.
ISA and ALNS-500 have close performance in terms of the observation profit.
Specifically, ALNS-500 is fractionally superior to ISA in the scenario with 500 targets, while ISA performs better in the scenarios with more observation targets. Meanwhile, ALNS-500 takes \blue{}a longer computational time than ISA.\@ 
\blue{}BDP-ILS consumes the longest computation time because 
	the observation start time has to be iteratively calculated for energy constraints evaluation.
\blue{}Relying on the fast insertion algorithm combined with a strategy to determine the observation start time, ISA can obtain scheduling results with high  efficiency. 
Overall, these results suggest that ISA outperforms \red{}ALNS, GA and BDP-ILS, which verifies the effectiveness of the proposed algorithm.

\subsubsection{Sensitivity analysis}

\begin{figure*}[htbp]
	\begin{center}
		\includegraphics[width=0.75\textwidth]{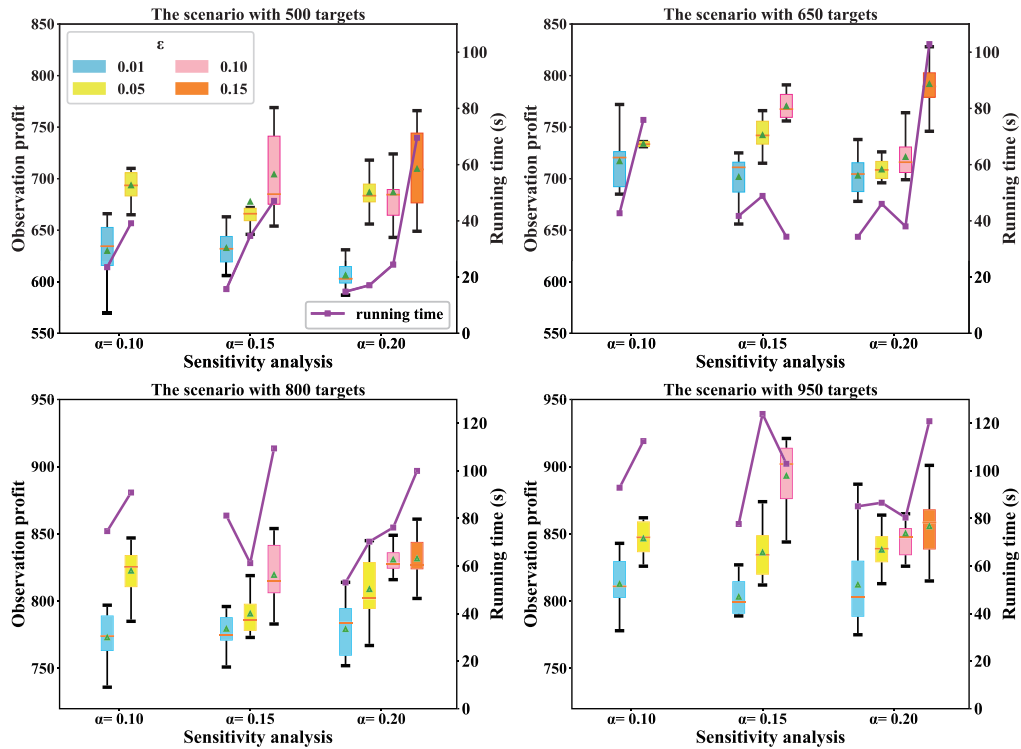}
		\caption{{\bfseries} Simulation results of sensitivity analysis with different number of targets. }\label{figEpsilon}
	\end{center}
\end{figure*}

The confidence interval and the confidence level of the sample approximation method
may impact the results of the observation profit and program running time. To test the influence of these factors, we adjust $1 - \alpha$ with 0.90, 0.85 and 0.80, $1 - \epsilon$ with 0.99, 0.95, 0.90 and 0.85. Notice that $1 - \epsilon$ $>$ $1 - \alpha$ should be maintained in line with the sample approximation method. $E_{m}$ and $M_{m}$ are fixed as 80 kJ and 7500 MB.
For each combination of $\alpha$ and $\epsilon$ in scenarios with 500, 650, 800 or 950 targets, 10 runs are conducted. Therefore we entirely have 9 $\times$ 4 $\times$ 10 = 360 runs to analyze the sensitivity of the ISA based heuristic.

\red{}The simulation results of different parameter combinations are reported in Fig.~\ref{figEpsilon}.
	Notably, parameter $\epsilon$ has a significant impact on the performance of the proposed algorithm. 
	It is apparent from each subplot in Fig.~\ref{figEpsilon} that the observation profit improves with the increase of $\epsilon$.
	Theoretically, a larger value of $\epsilon$ means a lower confidence level, resulting in the acceptance of a less conservative solution, which complies with the simulation results. 
	In summary, the ISA based heuristic overall performs well in different settings of the confidence level of the sample approximation method, indicating the broad feasibility of the proposed algorithm.

\section{Conclusions}
\label{Sec:Concl}

The multiple AEOSs scheduling problem under cloud coverage uncertainty is addressed in this work. 
We aim to maximize the total observation profit under a predefined confidence level. The constraint satisfaction model is constructed, where the energy \& memory constraints and attitude maneuverability constraint are taken into consideration. The CCP method is further introduced to describe the cloud coverage uncertainty. However, it is difficult to calculate the observation profit in the CCP model. The sample approximation method is then adopted to reformulate the chance constraint and calculate the target profit, where the feasibility is guaranteed within a predefined confidence level. 
Subsequently, the optimization subproblem based on the concept of time slack is defined to determine the~\red{}observation start time. Finally, the \red{}ISA based heuristic based upon the fast insertion algorithm is developed to iteratively optimize observation profit under cloud coverage uncertainty. 
Computational results show that the \red{}ISA based heuristic could efficiently obtains solution for multi-AEOS scheduling with cloud coverage uncertainty. \red{}Compared to ALNS, GA and BDP-ILS in AEOSs scheduling under cloud coverage uncertainty, the proposed heuristic could obtain better scheduling results utilizing a shorter time, which verifies the efficiency and effectiveness of our algorithm.

Future research on AEOSs scheduling considering cloud coverage uncertainty may be oriented in two directions.
The objective function in this study is to maximize the total confidence observation profit, while the observation angle of AEOS could also affect the observation missions. Therefore, multi-objective optimization method could be taken into consideration.   In addition, the onboard autonomous scheduling for multiple AEOSs is more practical in dealing with various uncertainties. This asserts a high claim for real-time scheduling, which deserves to be further studied.
\section*{Acknowledgment}
\red{}Our deepest gratitude goes to the anonymous reviewers for their
careful work and thoughtful suggestions that have improved this paper
substantially. We also thank Guansheng Peng for enthusiastic discussions.

\ifCLASSOPTIONcaptionsoff
  \newpage
\fi



%

\bibliographystyle{IEEEtran}
\bibliography{SatScheduling}

 \newcommand{\noop}[1]{}
\begin{thebibliography}{10}
\providecommand{\url}[1]{#1}
\csname url@samestyle\endcsname
\providecommand{\newblock}{\relax}
\providecommand{\bibinfo}[2]{#2}
\providecommand{\BIBentrySTDinterwordspacing}{\spaceskip=0pt\relax}
\providecommand{\BIBentryALTinterwordstretchfactor}{4}
\providecommand{\BIBentryALTinterwordspacing}{\spaceskip=\fontdimen2\font plus
\BIBentryALTinterwordstretchfactor\fontdimen3\font minus
  \fontdimen4\font\relax}
\providecommand{\BIBforeignlanguage}[2]{{%
\expandafter\ifx\csname l@#1\endcsname\relax
\typeout{** WARNING: IEEEtran.bst: No hyphenation pattern has been}%
\typeout{** loaded for the language `#1'. Using the pattern for}%
\typeout{** the default language instead.}%
\else
\language=\csname l@#1\endcsname
\fi
#2}}
\providecommand{\BIBdecl}{\relax}
\BIBdecl

\bibitem{liu2017adaptive}
X.~Liu, G.~Laporte, Y.~Chen, and R.~He, ``An adaptive large neighborhood search
  metaheuristic for agile satellite scheduling with time-dependent transition
  time,'' \emph{Computers \& Operations Research}, vol.~86, pp. 41--53, 2017.

\bibitem{chen2019mixed}
X.~Chen, G.~Reinelt, G.~Dai, and A.~Spitz, ``A mixed integer linear programming
  model for multi-satellite scheduling,'' \emph{European Journal of Operational
  Research}, vol. 275, no.~2, pp. 694--707, 2019.

\bibitem{wang2019onboard}
X.~Wang, C.~Han, P.~Yang, and X.~Sun, ``Onboard satellite visibility prediction
  using metamodeling based framework,'' \emph{Aerospace Science and
  Technology}, vol.~94, p. 105377, 2019.

\bibitem{gu2019kriging}
Y.~Gu, C.~Han, and X.~Wang, ``A kriging based framework for rapid
  satellite-to-site visibility determination,'' in \emph{2019 IEEE 10th
  International Conference on Mechanical and Aerospace Engineering
  (ICMAE)}.\hskip 1em plus 0.5em minus 0.4em\relax IEEE, 2019, pp. 262--267.

\bibitem{HE201812}
L.~He, X.~Liu, G.~Laporte, Y.~Chen, and Y.~Chen, ``An improved adaptive large
  neighborhood search algorithm for multiple agile satellites scheduling,''
  \emph{Computers \& Operations Research}, vol. 100, pp. 12--25, 2018.

\bibitem{LemaitreVerfaillie-269}
M.~Lemaître, G.~Verfaillie, F.~Jouhaud, J.-M. Lachiver, and N.~Bataille,
  ``Selecting and scheduling observations of agile satellites,''
  \emph{Aerospace Science and Technology}, vol.~6, no.~5, pp. 367--381, 2002.

\bibitem{GabrelMoulet-258}
V.~Gabrel, A.~Moulet, C.~Murat, and V.~T. Paschos, ``A new single model and
  derived algorithms for the satellite shot planning problem using graph theory
  concepts,'' \emph{Annals of Operations Research}, vol.~69, pp. 115--134,
  1997.

\bibitem{GabrelVanderpooten-61}
V.~Gabrel and D.~Vanderpooten, ``Enumeration and interactive selection of
  efficient paths in a multiple criteria graph for scheduling an {E}arth
  observing satellite,'' \emph{European Journal of Operational Research}, vol.
  139, no.~3, pp. 533--542, 2002.

\bibitem{WolfeSorensen-272}
W.~J. Wolfe and S.~E. Sorensen, ``Three scheduling algorithms applied to the
  {E}arth observing systems domain,'' \emph{Management Science}, vol.~46,
  no.~1, pp. 148--166, 2000.

\bibitem{Benoist2004Upper}
T.~Benoist and B.~Rottembourg, ``Upper bounds for revenue maximization in a
  satellite scheduling problem,'' \emph{Quarterly Journal of the Belgian French
  \& Italian Operations Research Societies}, vol.~2, no.~3, pp. 235--249, 2004.

\bibitem{LinLiao-60}
W.~C. Lin, D.~Y. Liao, C.~Y. Liu, and Y.~Y. Lee, ``Daily imaging scheduling of
  an {E}arth observation satellite,'' \emph{IEEE Transactions on Systems, Man,
  and Cybernetics - Part A: Systems and Humans}, vol.~35, no.~2, pp. 213--223,
  2005.

\bibitem{Marinelli2011A}
F.~Marinelli, S.~Nocella, F.~Rossi, and S.~Smriglio, ``A lagrangian heuristic
  for satellite range scheduling with resource constraints,'' \emph{Computers
  \& Operations Research}, vol.~38, no.~11, pp. 1572--1583, 2011.

\bibitem{wu2013two}
G.~Wu, J.~Liu, M.~Ma, and D.~Qiu, ``A two-phase scheduling method with the
  consideration of task clustering for {E}arth observing satellites,''
  \emph{Computers \& Operations Research}, vol.~40, no.~7, pp. 1884--1894,
  2013.

\bibitem{wu2015coordinated}
G.~Wu, W.~Pedrycz, H.~Li, M.~Ma, and J.~Liu, ``Coordinated planning of
  heterogeneous {E}arth observation resources,'' \emph{IEEE Transactions on
  Systems, Man, and Cybernetics: Systems}, vol.~46, no.~1, pp. 109--125, 2015.

\bibitem{wu2017satellite}
G.~Wu, H.~Wang, W.~Pedrycz, H.~Li, and L.~Wang, ``Satellite observation
  scheduling with a novel adaptive simulated annealing algorithm and a dynamic
  task clustering strategy,'' \emph{Computers \& Industrial Engineering}, vol.
  113, pp. 576--588, 2017.

\bibitem{wang2018fixed}
X.~Wang, R.~Leus, and C.~Han, ``Fixed interval scheduling of multiple {E}arth
  observation satellites with multiple observations,'' in \emph{2018 9th
  International Conference on Mechanical and Aerospace Engineering
  (ICMAE)}.\hskip 1em plus 0.5em minus 0.4em\relax IEEE, 2018, pp. 28--33.

\bibitem{hu2019branch}
X.~Hu, W.~Zhu, B.~An, P.~Jin, and W.~Xia, ``A branch and price algorithm for
  {EOS} constellation imaging and downloading integrated scheduling problem,''
  \emph{Computers \& Operations Research}, vol. 104, pp. 74--89, 2019.

\bibitem{wang2020agile}
X.~Wang, G.~Wu, L.~Xing, and W.~Pedrycz, ``Agile earth observation satellite
  scheduling over 20 years: formulations, methods and future directions,''
  \emph{IEEE Systems Journal}, 2020, doi: 10.1109/JSYST.2020.2997050.

\bibitem{Beaumet2011Feasibility}
G.~Beaumet, G.~Verfaillie, and M.~C. Charmeau, ``Feasibility of autonomous
  decision making on board an agile {E}arth-observing satellite,''
  \emph{Computational Intelligence}, vol.~27, no.~1, pp. 123--139, 2011.

\bibitem{liu2016time}
S.~Liu, Y.~Chen, L.~Xing, and X.~Guo, ``Time-dependent autonomous task planning
  of agile imaging satellites,'' \emph{Journal of Intelligent \& Fuzzy
  Systems}, vol.~31, no.~3, pp. 1365--1375, 2016.

\bibitem{she2018onboard}
Y.~She, S.~Li, and Y.~Zhao, ``Onboard mission planning for agile satellite
  using modified mixed-integer linear programming,'' \emph{Aerospace Science
  and Technology}, vol.~72, pp. 204--216, 2018.

\bibitem{wang2016scheduling}
X.-W. Wang, Z.~Chen, and C.~Han, ``Scheduling for single agile satellite,
  redundant targets problem using complex networks theory,'' \emph{Chaos,
  Solitons \& Fractals}, vol.~83, pp. 125--132, 2016.

\bibitem{wang2019scheduling}
X.~Wang, C.~Han, R.~Zhang, and Y.~Gu, ``Scheduling multiple agile earth
  observation satellites for oversubscribed targets using complex networks
  theory,'' \emph{IEEE Access}, vol.~7, pp. 110\,605--110\,615, 2019.

\bibitem{cho2018optimization}
D.-H. Cho, J.-H. Kim, H.-L. Choi, and J.~Ahn, ``Optimization-based scheduling
  method for agile {E}arth-observing satellite constellation,'' \emph{Journal
  of Aerospace Information Systems}, vol.~15, no.~11, pp. 611--626, 2018.

\bibitem{li2018preference}
L.~Li, H.~Chen, J.~Li, N.~Jing, and M.~Emmerich, ``Preference-based
  evolutionary many-objective optimization for agile satellite mission
  planning,'' \emph{IEEE Access}, vol.~6, pp. 40\,963--40\,978, 2018.

\bibitem{du2018area}
B.~Du, S.~Li, Y.~She, W.~Li, H.~Liao, and H.~Wang, ``Area targets observation
  mission planning of agile satellite considering the drift angle constraint,''
  \emph{Journal of Astronomical Telescopes, Instruments, and Systems}, vol.~4,
  no.~4, p. 047002, 2018.

\bibitem{Guansheng2020An}
G.~Peng, G.~Song, L.~Xing, A.~Gunawan, and P.~Vansteenwegen, ``An exact
  algorithm for agile {E}arth observation satellite scheduling with
  time-dependent profits,'' \emph{Computers \& Operations Research}, vol. 120,
  p. 104946, 2020.

\bibitem{wang2014dynamic}
J.~Wang, X.~Zhu, D.~Qiu, and L.~T. Yang, ``Dynamic scheduling for emergency
  tasks on distributed imaging satellites with task merging,'' \emph{IEEE
  Transactions on Parallel and Distributed Systems}, vol.~25, no.~9, pp.
  2275--2285, 2014.

\bibitem{WangZhu-21}
J.~Wang, X.~Zhu, L.~T. Yang, J.~Zhu, and M.~Ma, ``Towards dynamic real-time
  scheduling for multiple {E}arth observation satellites,'' \emph{Journal of
  Computer and System Sciences}, vol.~81, no.~1, pp. 110--124, 2015.

\bibitem{du2018new}
B.~Du and S.~Li, ``A new multi-satellite autonomous mission allocation and
  planning method,'' \emph{Acta Astronautica}, 2018.

\bibitem{ntagiou2018ant}
E.~V. Ntagiou, C.~Iacopino, N.~Policella, R.~Armellin, and A.~Donati,
  ``Ant-based mission planning: Two examples,'' in \emph{2018 SpaceOps
  Conference}, 2018, p. 2498.

\bibitem{he2019hierarchical}
L.~He, X.-L. Liu, Y.-W. Chen, L.-N. Xing, and K.~Liu, ``Hierarchical scheduling
  for real-time agile satellite task scheduling in a dynamic environment,''
  \emph{Advances in Space Research}, vol.~63, no.~2, pp. 897--912, 2019.

\bibitem{globus2004comparison}
A.~Globus, J.~Crawford, J.~Lohn, and A.~Pryor, ``A comparison of techniques for
  scheduling {E}arth observing satellites,'' in \emph{Association for the
  Advancement of Artificial Intelligence}, 2004, pp. 836--843.

\bibitem{wang2019robust}
X.~Wang, G.~Song, R.~Leus, and C.~Han, ``Robust earth observation satellite
  scheduling with uncertainty of cloud coverage,'' \emph{IEEE Transactions on
  Aerospace and Electronic Systems}, 2019.

\bibitem{JU20081196}
J.~Ju and D.~P. Roy, ``The availability of cloud-free {L}andsat {ETM}+ data
  over the conterminous {U}nited {S}tates and globally,'' \emph{Remote Sensing
  of Environment}, vol. 112, no.~3, pp. 1196--1211, 2008.

\bibitem{he2013research}
M.~He and R.~He, ``Research on agile imaging satellites scheduling techniques
  with the consideration of cloud cover,'' \emph{Sci Technol Eng}, vol.~13,
  no.~28, pp. 8373--8379, 2013.

\bibitem{liao2007imaging}
D.-Y. Liao and Y.-T. Yang, ``Imaging order scheduling of an {E}arth observation
  satellite,'' \emph{IEEE Transactions on Systems, Man, and Cybernetics, Part C
  (Applications and Reviews)}, vol.~37, no.~5, pp. 794--802, 2007.

\bibitem{WangDemeulemeester-4}
J.~Wang, E.~Demeulemeester, and D.~Qiu, ``A pure proactive scheduling algorithm
  for multiple {E}arth observation satellites under uncertainties of clouds,''
  \emph{Computers \& Operations Research}, vol.~74, pp. 1--13, 2016.

\bibitem{valicka2019mixed}
C.~G. Valicka, D.~Garcia, A.~Staid, J.-P. Watson, G.~Hackebeil, S.~Rathinam,
  and L.~Ntaimo, ``Mixed-integer programming models for optimal constellation
  scheduling given cloud cover uncertainty,'' \emph{European Journal of
  Operational Research}, vol. 275, no.~2, pp. 431--445, 2019.

\bibitem{8573152}
J.~Wang, E.~Demeulemeester, X.~Hu, D.~Qiu, and J.~Liu, ``Exact and heuristic
  scheduling algorithms for multiple {E}arth observation satellites under
  uncertainties of clouds,'' \emph{IEEE Systems Journal}, pp. 1--12, 2018.

\bibitem{wang2020expectation}
J.~Wang, E.~Demeulemeester, X.~Hu, and G.~Wu, ``Expectation and {SAA} models
  and algorithms for scheduling of multiple {E}arth observation satellites
  under the impact of clouds,'' \emph{IEEE Systems Journal}, 2020, doi:
  10.1109/JSYST.2019.2961236.

\bibitem{he2016cloud}
L.~He, X.~Liu, L.~Xing, and Y.~Chen, ``Cloud avoidance scheduling algorithm for
  agile optical satellites,'' \emph{Journal of Computational and Theoretical
  Nanoscience}, vol.~13, no.~6, pp. 3691--3705, 2016.

\bibitem{Bensana1999Dealing}
E.~Bensana, G.~Verfaillie, C.~Michelon-Edery, and N.~Bataille, ``Dealing with
  uncertainty when managing an {E}arth observation satellite,'' \emph{European
  Space Agency-Publications-ESA SP}, vol. 440, pp. 205--210, 1999.

\bibitem{liu2009some}
B.~Liu, ``Some research problems in uncertainty theory,'' \emph{Journal of
  Uncertain Systems}, vol.~3, no.~1, pp. 3--10, 2009.

\bibitem{peng2019agile}
G.~Peng, R.~Dewil, C.~Verbeeck, A.~Gunawan, L.~Xing, and P.~Vansteenwegen,
  ``Agile {E}arth observation satellite scheduling: An orienteering problem
  with time-dependent profits and travel times,'' \emph{Computers \& Operations
  Research}, vol. 111, pp. 84--98, 2019.

\bibitem{luedtke2008sample}
J.~Luedtke and S.~Ahmed, ``A sample approximation approach for optimization
  with probabilistic constraints,'' \emph{SIAM Journal on Optimization},
  vol.~19, no.~2, pp. 674--699, 2008.

\bibitem{metropolis1949monte}
N.~Metropolis and S.~Ulam, ``The monte carlo method,'' \emph{Journal of the
  American statistical association}, vol.~44, no. 247, pp. 335--341, 1949.

\bibitem{ruszczynski2002probabilistic}
A.~Ruszczy{\'n}ski, ``Probabilistic programming with discrete distributions and
  precedence constrained knapsack polyhedra,'' \emph{Mathematical Programming},
  vol.~93, no.~2, pp. 195--215, 2002.

\bibitem{danielsson1980euclidean}
P.-E. Danielsson, ``Euclidean distance mapping,'' \emph{Computer Graphics and
  Image Processing}, vol.~14, no.~3, pp. 227--248, 1980.

\bibitem{boggs1995sequential}
P.~T. Boggs and J.~W. Tolle, ``Sequential quadratic programming,'' \emph{Acta
  Numerica}, vol.~4, pp. 1--51, 1995.

\bibitem{wang2018multi}
R.~Wang, S.~Lai, G.~Wu, L.~Xing, L.~Wang, and H.~Ishibuchi, ``Multi-clustering
  via evolutionary multi-objective optimization,'' \emph{Information Sciences},
  vol. 450, pp. 128--140, 2018.

\bibitem{zhang2019large}
J.~Zhang, L.~Wang, and L.~Xing, ``Large-scale medical examination scheduling
  technology based on intelligent optimization,'' \emph{Journal of
  Combinatorial Optimization}, vol.~37, no.~1, pp. 385--404, 2019.

\bibitem{xiang2019comprehensive}
S.~Xiang, L.~Xing, L.~Wang, and K.~Zou, ``Comprehensive learning
  pigeon-inspired optimization with tabu list,'' \emph{Science China
  Information Sciences}, vol.~62, no.~7, p. 70208, 2019.

\bibitem{kirkpatrick1983optimization}
S.~Kirkpatrick, C.~D. Gelatt, and M.~P. Vecchi, ``Optimization by simulated
  annealing,'' \emph{Science}, vol. 220, no. 4598, pp. 671--680, 1983.

\bibitem{kirkpatrick1984optimization}
S.~Kirkpatrick, ``Optimization by simulated annealing: Quantitative studies,''
  \emph{Journal of Statistical Physics}, vol.~34, no. 5-6, pp. 975--986, 1984.

\bibitem{han2018scheduling}
C.~Han, X.~Wang, G.~Song, and R.~Leus, ``Scheduling multiple agile {E}arth
  observation satellites with multiple observations,'' \emph{arXiv preprint
  arXiv:1812.00203}, 2018.

\end{thebibliography}

%




\end{document}